**Title**

# Accurate Inverse Process Optimization Framework in Laser Directed Energy Deposition

**Author list**

Xiao Shang[1], Evelyn Li[1], Ajay Talbot[1], Haitao Wen[1], Tianyi Lyu[1], Jiahui Zhang[1], Yu Zou[1] *

**Affiliations**

[1]Department of Materials Science and Engineering, University of Toronto, Toronto, ON,

Canada, M5S 3E4

*Corresponding author. Email: mse.zou@utoronto.ca

**Abstract**

In additive manufacturing (AM), particularly for laser-based metal AM, process optimization is crucial to the quality of products and the efficiency of production. The identification of optimal process parameters out of a vast parameter space, however, is a daunting task. Despite advances in simulations, the process optimization for specific materials and geometries is developed through a sequential and time-consuming trial-and-error approach, which often lacks the versatility to address multiple optimization objectives. Machine learning (ML) provides a powerful tool to accelerate the optimization process, but most current studies focus on simple single-track prints, which hardly translate to manufacturing 3D components for engineering applications. In this study, we develop an *A*ccurate *I*nverse process optimization framework in laser *D*irected *E*nergy *D*eposition (AIDED), based on machine learning models and a genetic algorithm, to aid process optimization in laser DED processes. Using the AIDED, we demonstrate the following: (i) Accurately predict single-track ($R^2$ score 0.995), multi-track ($R^2$ score 0.969), and multi-layer (1.07% and 10.75% error in width and height, respectively) cross-sectional melt pool geometries directly from process parameters; (ii) Determine appropriate hatch spacing and layer thickness for fabricating fully dense (density > 99.9%) multi-track and multi-layer prints; (iii) Inversely identify optimal process parameters directly from customizable application objectives within 1-3 hours. We also validate the effectiveness of the AIDED experimentally by achieving two exemplary targets – fast print speed and fine print resolution. Furthermore, we show the high transferability of the framework from stainless steel to pure nickel. With such transferability in AIDED, we pave a new way for "aiding" the process optimization in the laser-based AM processes that is applicable to a wide range of materials.

## 1. Introduction

Additive manufacturing (AM), especially laser-based metal AM or 3D printing, has brought to the modern industry revolutionary potentials due to its capability of manufacturing components with complex shapes [1-4]. Among the most widely used metal AM technologies, laser directed energy deposition (L-DED) stands out as a preferred method [5] for multi-material printing [6-8], parts reparation [9, 10], and large near-net-shape component manufacturing [11, 12]. In the L-DED processes, process parameters such as laser power ($P$), print speed ($v$), and powder feed rate ($\dot{m}$) control the complex thermal history during the printing process. As a result, the quality and properties of the printed components, such as their porosity [13, 14], dimensional accuracy [14-16], and mechanical properties [17-19], depend on the process parameters. Thus, one major challenge to the wider adaptation of L-DED, and more broadly AM, is the selection of optimal process parameters [5, 14, 16, 20].

Consistent efforts have been made to process optimization, among which offline methods using empirical, analytical, and simulation-based techniques, as well as online control methods, have been widely studied. Offline methods aim to establish the process-structure-property (P-S-P) relations to guide the selection of process parameters asynchronous to the printing process. For example, through the systematic design of experiments, processing maps are produced to provide a guideline to determine process parameters [14, 16, 21]. By analytically modeling the energy input as a result of process parameters, preferred parameters can be identified to optimize the printing quality [15, 22-24]. Using numerical simulations such as finite element analysis (FEA) and computational fluid dynamics (CFD), more accurate physic-informed models are established, allowing for relatively precise estimations of the melt-to-resolidify process during printing [13, 25-27]. In contrast, online methods adjust process parameters during the printing process according to certain feedback mechanisms. Leveraging a wide range of sensors such as visible light and infrared cameras, melt pool signatures are captured to provide feedback to update process parameters in real time [28-33]. Nevertheless, these approaches show disadvantages that hurdle their wider applications. For example, in offline approaches, empirical methods require vast time and monetary investments into experiments and are often non-transferrable across metal 3D printers and materials systems. Despite offering rapid calculation capabilities and thus being time-saving, analytical methods often sacrifice accuracy. Although simulation methods are accurate, the amount of time and computational resources required render them less preferable in real-time industrial applications. While online approaches are typically rapid and accurate, the cost for complex hardware setup makes it less attractive for industries to adapt.

Over the past decade, machine learning (ML) and statistical approaches have received increasing attention for AM, due to their capabilities to learn highly nonlinear P-S-P relations, swift inference time, and good transferability [34-37]. For the process optimization in AM, ML has been used for predicting the geometrical characteristics of the melt pool, a key indicator for the quality of additively manufactured parts [38-41]. However, to date, limited work has been done for L-DED. Akbari et al. created a comprehensive ML model for estimating basic melt pool geometries including the width, depth, and length, but less than 3% of the dataset used in training their model is for L-DED [35]. Feenstra et al. leveraged neural networks (NN) to predict melt pool geometries and produced processing maps for rationalise processing windows in L-DED [42]. However, in these studies, only single tracks (exemplary prints in Figure 1 (a). All prints in Figure 1 are prepared using a commonly used L-DED system in our laboratory) were considered, which is insufficient to reflect the melt pool geometries in the more crucial multi tracks and multi layers (exemplary prints in Figure 1 (d) and (f))). Moreover, only basic melt pool geometrical features such as the width ($W$), height ($H$), and depth ($D$) are predicted in the previous studies (exemplary single-track melt pools in Figure 1 (b) and (c)), compromising the accuracy of the predictions. Furthermore, with the predicted geometrical features, researchers are limited to reconstructing the full melt pool shapes using simple models, such as quadratic models stacked in an overlapping way [14, 43, 44], yielding significant estimation errors not only for the single track case but accumulatively for the multi-track and multi-layer cases. As evident in Figure 1 (b) and (c), both melt pools are similar in $W$, $H$, and $D$, yet their geometrical shapes are largely different. Meanwhile, none of the melt pool contour is accurately captured by using a quadratic model (dashed white lines). The multi-track and multi-layer prints in Figure 1 (e) and (g) give a comparison between real-life melt pool arrangements and common estimations, which are created by overlapping single-track melt pools. Compared with the commonly used estimations (dashed white lines), in real-life (solid black lines) the melt pools stack on one another, consequently changing their geometrical shapes compared with the single-track prints.

Moreover, few studies to date demonstrate the capability of inversely identifying optimal process parameters directly from target applications. Shi et al. propose a data-driven method for determining the process parameters in AM, but it only focuses on lattice materials [45]. Wang et al. describe another data-driven approach targeting the inverse identification of robust process parameters for optimising the microstructure, but the method is not validated with experimental data, nor is the transferability shown for different material systems and optimization applications [46]. Similar work also includes [47], where a multi-objective optimization approach is demonstrated, focusing on the fused filament fabrication AM process for polymers. Therefore, it is highly desirable to develop an approach that can identify optimal process parameters inversely from application objectives. Such approach should be robust, validated experimentally, and readily transferrable to various materials and applications.

In this study, we propose an *A*ccurate *I*nverse process parameter optimization framework for L-*DED* (AIDED) empowered by ML and a genetic algorithm, and describe in detail how it is progressively developed from single-track to multi-track, and finally to multi-layer prints. The widely used materials in L-DED, grade 316L stainless steel and PH17-4 steel, are used as the exemplary material system. The good transferability of the framework is demonstrated using pure nickel. Lastly, we showcase inverse process optimization to search for the set of process parameters that yields optimal print speed and print resolution, before validating the results with experiments.

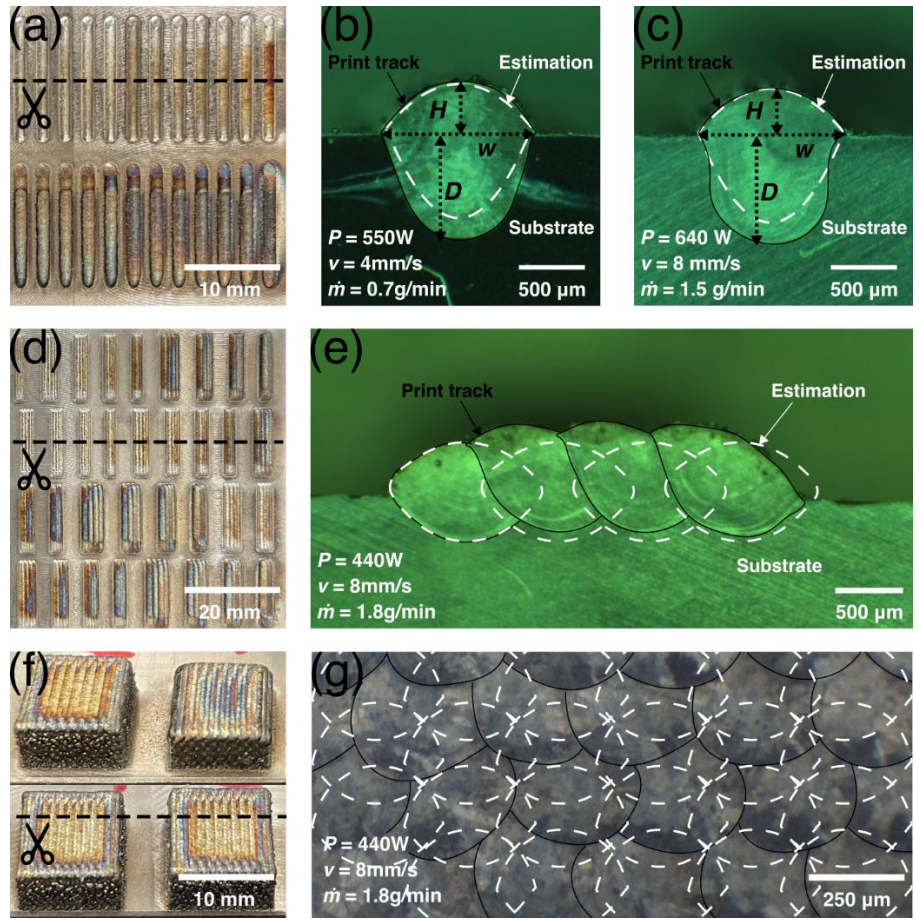

**Figure 1.** Sample single-track, multi-track, and multi-layer L-DED prints and their cross-sectional geometries. (a) Top view of typical single-track prints. (b) and (c) Optical microscope (OM) images showing cross-sectional melt pool arrangements of two single-track prints. Real-life melt pool contours and corresponding quadratic estimations are marked with solid black lines and dashed white lines, respective. The basic melt pool measurements are indicated with dashed black arrows. (d) Top view of typical multi-track prints. (e) OM image showing cross-sectional melt pool arrangements of a multi-track print and the comparison between real-life contours and common estimations. (f) Trimetric and top views of typical multi-layer prints. (g) OM image showing cross-sectional melt pool arrangements of a multi-layer print and the comparison between real-life contours and common estimations. In (e) and (g), the solid black lines are real-life melt pool contours, and dashed white lines indicate commonly accepted estimations.

## 2. Materials and methods

### 2.1 L-DED platform and materials

In this work, all samples were manufactured using an in-house developed 3-axis L-DED system. The power source (IPG Photonics Corporation) is a fibre laser with a maximum power output of 1 kW and a wavelength of 1070 nm. The nominal beam size of the laser is 1 mm. A coaxial printing nozzle (COAX14 v5, Fraunhofer USA, Inc.) was used as the printing nozzle, and argon was used as the purge gas to prevent oxidation. At the start of each new print, the nozzle was positioned 7.5 mm above the substrate. During printing, this distance was maintained by setting the appropriate layer thickness (*t*). A mass flow meter and PLC-controlled powder feeder (PF 2/2, GTV GmbH) were used to accurately deliver the desired amount of powder. During the printing process, the laser power, scan speed, as well as powder federate were controlled by the G-code programming language. The powder feed rate (g/min) was normalized as the disc rotation speed in revolutions per minute (rpm). A linear relationship between the powder feed rate and disc rotation speed was measured as the following equation:

$$feedrate = disc\ rotation\ speed \times 3.5 \tag{1}$$

The stainless steel 316 and PH17-4 powders used in this work were acquired from EOS GmbH and the pure nickel (Ni) powder was acquired from MSE Supplies LLC. Both powders have a size distribution of 15-45 μm. Their compositions are listed in Table 1. Printing of the 316L and PH17-4 steel sample were done on ½" thick 316L stainless steel boards and the Ni samples were done on ½" low alloy steel boards. Before printing, the surfaces of the printing boards were ground with grit 60 sandpapers for repeatability.

**Table 1**

Chemical composition of the metal powders used in this work.

| Material | Fe | Cr | Ni | Mo | C | N | Cu | Nb + Ta |
|---|---|---|---|---|---|---|---|---|
| **SS 316L** | Bal. | 17.00-19.00 | 13.00-15.00 | 2.25-3.00 | 0-0.03 | 0-0.10 | NA | NA |
| **PH17-4** | Bal. | 17.00 | 4.50 | NA | <0.07 | NA | 4.00 | 0.3 |
| **Ni** | NA | NA | 99.9% | NA | NA | NA | NA | NA |

**2.2. Printing of samples and characterization of melt pools**

To generate a comprehensive dataset for training the ML models, we selected process parameters covering a wide processing window as illustrated in Figure 2 (a) and Table 2. For single-track prints, the following parameters were selected: laser powers ranging 280-640 W with a step size of 10 W, scan speed ranging 4-12 mm/s with a step size of 4 mm/s, and powder feed rate ranging 0.2-0.8 rpm with a step size of 0.2 rpm This way, a wide range of linear energy density (LED) was covered, yielding a single-track dataset with 444 diverse melt pool geometries (Figure 2(b)). Here the LED is defined as:

$$LED = \frac{P}{v \times \dot{m}} \qquad (2)$$

Where *P* is the laser powder in W, *v* the scan speed in mm/s, and $\dot{m}$ the mass flow rate in g/min. A top view of the complete print board with all single tracks is shown in Figure S1 (a).

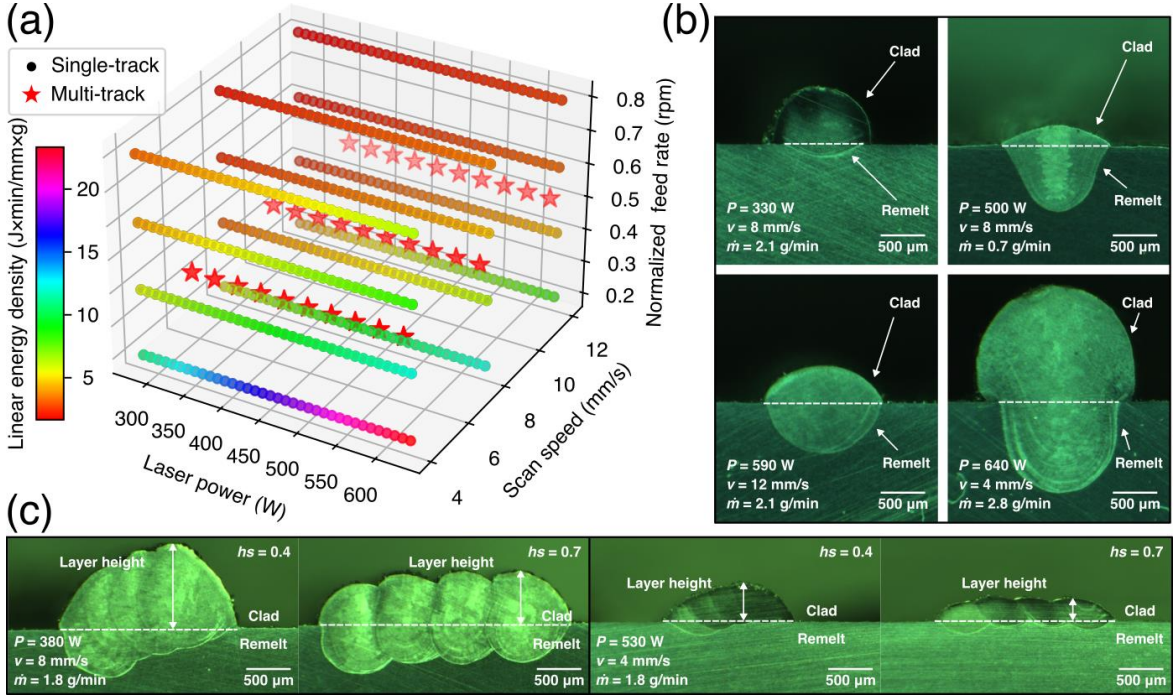

**Figure 2.** Process parameter design space and exemplary samples of the single-track and multi-track prints. (a) Process parameter design space for single-track and multi-track prints. (b) Sample microscopic images of single-track melt pool cross-sections under various process parameters. The clad and remelt areas are marked with white arrows. (c) Sample microscopic images of multi-track melt pool cross-sections under various process parameters. The clad and remelt areas are marked with white arrows.

**Table 2**

Process parameter windows used for data generation

| Material | Laser power, $P$, (w) | Scan speed, $v$, (mm/s) | Powder feed rate, $\dot{m}$, (g/min) | Linear energy density (j×min/mm×g) |
|---|---|---|---|---|
| **316L single-track** | 280-640 | 4-12 | 0.7-2.8 | 8.3-228.6 |
| **316L multi-track** | 280-640 | 4-12 | 1.8 | 13.3-91.4 |
| **Nickel** | 270-640 | 4-12 | 0.7-2.8 | 13.9-207.14 |

For multi-track prints, we used laser powers ranging 280-640 W with a step size of 30 W, and scan speed ranging 4-12 mm/s with a step size of 4 mm/s. For the powder feed rate, 0.5 rpm was used to provide a meaningful representation of a middle-range value compared with the single-track dataset. Furthermore, a new process parameter, hatch spacing ($hs$), was introduced into the multi-track case (Figure 3 (a)). $Hs$ was used to measure how far the centre of each single track is from each other when being deposited and is a relative value to the melt pool width. In this work, $hs$ values from 0.4-0.7, with a step size of 0.1 were selected. The multi-track printing path used in this work is shown in Figure 3 (a), where the print direction was inversed in-between neighbouring tracks. In total, 112 multi-track prints with diverse melt pool geometries and geometries were generated for the multi-track dataset (Figure 2 (c)). A top view of the complete print board with all multi-tracks is shown in Figure S1 (b).

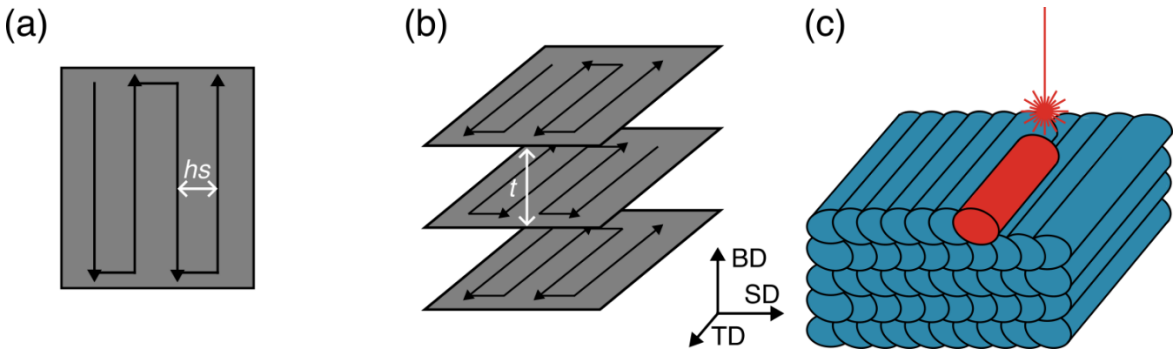

**Figure 3.** Schematic illustrations of the printing paths used in this work. (a) In-plane multi-track scanning path. The printing direction changes for each neighbouring track. The hatch

spacing ($hs$) is marked with a white arrow. (b) Mult-layer scanning path. The printing path reverses for each neighbouring layer. The layer thickness ($t$) is marked with a white arrow. (c) A 3D schematic of multi-layer print with the three printing directions, i.e., scan direction (SD), transverse direction (TD), and build direction (BD), marked.

Multi-layer prints were produced as validators for the proposed process parameter optimization framework. Thus, their printing parameters were determined solely by the optimization results. Multi-layer prints using two materials, 316L stainless steel and PH17-4 steel, were printed to demonstrate the optimization results of this framework. PH17-4 was chosen because of its similarity in printability with 316L stainless steel, as well as its contrast in appearance with the latter after etching (Figure 4). This facilitates the demonstration of print resolution, one of the two optimization objectives in this work. The alternation of the two materials was controlled by G-code to time the amount and type of powder delivered during the printing process. Figure 3 (b) shows the printing path of multi-layer prints. In between the layers, the printing direction was inversed. The layer thickness ($t$) is the nozzle movement distance in height between layers. Figure 3(c) gives a schematic view of a multi-layer print, and the three printing directions, i.e., scan direction (SD), transverse direction (TD), and build direction (BD), are shown. A 5-second pause time was implemented between each print track to allow the temperature of a previous layer to drop below 80°C to reduce the influence of heat accumulation (Figure S3). For multi-material prints, this pause was set to 15 seconds for the powder stream to stabilize.

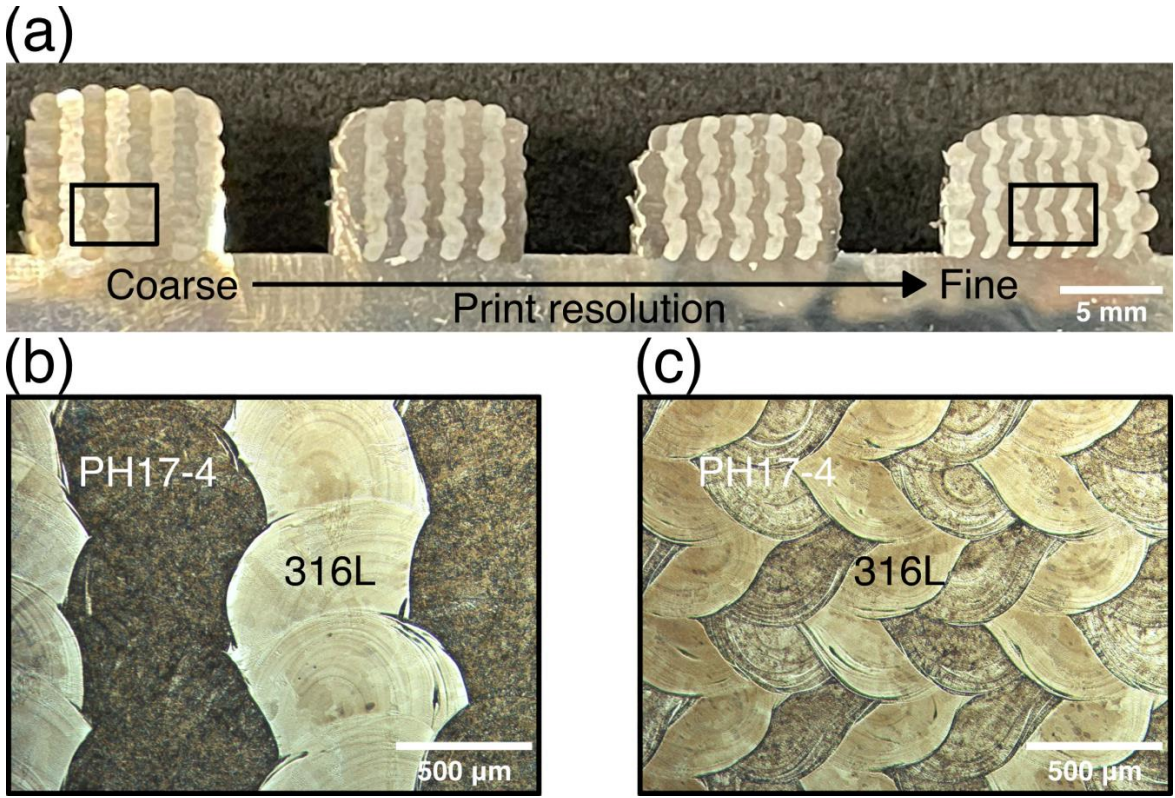

**Figure 4.** The OM images that illustrate print resolution using 316L stainless steel (light region) and PH17-4 steel (dark region) bi-material heterostructured cubic prints. (a) Cross

sections of four cubic prints with resolutions from coarse to fine. (b) and (c) Magnifcations of the views of the coarsest and finest prints.

To extract the melt pool cross-sections after printing, a series of sample preparation and characterization were performed. The prints were first machined into halves so the cross-sectional area could be accessed. The cross-sections were then ground and polished down to 0.1 µm polishing agent, before being etched with 50% concentration Carpenter solution (for 316L and Ph17-4 steels) or Adler solution (for Ni) to reveal the melt pool boundaries. The etched melt pools were then captured using a USB camera (Omron Sentech Co., LTD) mounted on our L-DED printer for automated image capture. A desktop optical microscope (Leica Microsystems GmbH) was then used to image the polished samples before the porosity was measured in ImageJ.

### 2.3. Machine learning methodology

### 2.3.1. Data preprocessing

The acquired melt pool images were first preprocessed for training the machine learning models (Figure 5 (a)). To process the large single-track dataset, we trained a YOLOv8 model [48] to detect and crop a bounding box around the melt pools. The exact melt pool contours were then extracted from the bounding boxes using the Segment Anything Model, a pre-trained image segmentation model that can "cut out" any object in any image [49]. The two-dimensional masks were then reduced via principal component analysis (PCA) to one-dimensional vectors with a size of 87 for training a regression model.

For multi-track prints, the first deposited melt pool shows the exact shape of its single-track counterpart, and the rest of the melt pools overlay with one another, consequently tilting at an angle $\alpha$ from the horizontal printing surface (Figure 5 (a)). Thus, to model the multi-track prints, we decomposed them into n equally spaced single-track melt pools spaced at $hs$. The first melt pool retains the shape of its single-track counterpart, while the rest are melt pools with the same shape but tilted at the angle $\alpha$. When equally spaced at $hs$, the complete multi-track were reconstructed. For all the multi-track prints in our dataset, their $\alpha$ was measured using imageJ for ML model training. In the final step, the multi-layer prints were reconstructed by stacking multi-track prints at the optimal layer printing thickness $t$ (Figure 5 (c)). The $t$ was determined as the distance between multi-track layers when no lack-of-fusion pores were found. As such, fully dense multi-layer cubes without lack-of-fusion pores were reconstructed.

### 2.3.2. Machine learning inference models

Two multi-layer perceptron (MLP) regressors were selected as the machine learning models for the prediction of single-track and multi-track melt pool geometries, respectively (Figure 5 (b)). As comparisons, two other types of machine learning models, namely linear regression (LR) and random forest (RF), were trained and tested side-by-side. In the case of single-track prints, $P$, $v$, and $\dot{m}$, herein referred to as the basic process parameters, were used as the input. The output is the PCA reduced-dimensional melt pool masks in the form of an 87-element vector. Through inverse PCA transformation, the vector can be reconstructed into a single-track melt pool mask, without losing the original information (Figure S3). For multi-track prints, we aim to predict the tilt angle ($\alpha$) resulting from the melt pool stacking. The $\alpha$ is a function of the basic process parameters, single-track melt pool geometry, as well as *hs*. Therefore, in the multi-track ML model, the basic process parameters, geometrical signatures (melt pool height (*H*) and width (*W*)) of the single-track melt pool, as well as *hs* were used as input. A correlation analysis was performed to determine the exact input being used (Figure S5), i.e. *v*, *H*, *W*, and *hs*.

The ML models were built using Scikit-learn [50] and TensorFlow [51], Python implementations for performing machine learning tasks. The models were trained on a desktop computer with a 16-core Intel[a] i7 CPU. Hyperparameter tuning and cross-validation were performed for both models to fine tune the models for optimal performance (Tables S1 and S2). The selected model architectures and parameters are described in Table S3. A train-test split ratio of 80-20 was used for the models, and all model evaluation metrics were calculated using corresponding stand-alone testing sets.

**2.4. Framework transferability**

In our framework, the single-track ML model predicts the melt pool size and geometry, which is a function of process parameters and the intrinsic thermal properties of a material. On the contrary, the multi-track ML model, the multi-layer reconstruction model, and the multi-objective optimization models mainly reflect the geometrical arrangements of the predicted single-track melt pools. Thus, to transfer the framework to a new material system, it is adequate to retrain solely a new single-track ML model for the new material. Nevertheless, it is a time-consuming process to collect a large dataset whenever a new material is used. To simplify such transferring process, we apply transfer learning techniques to the existing model trained with our comprehensive 316L stainless steel data. High-performing models were obtained with just a small portion of experimental data.

To demonstrate the framework's transferability, a separate single-track data set printed with pure Ni was generated, containing 71 single tracks within the printable process window (Table 2) of pure Ni. Only 56 images were used as the training set for retraining the model. When performing transfer learning, the first hidden layer was used as a pre-trained model and only the weights in the second hidden layer were updated through the training process (Table S3).

**2.5. Multi-objective optimization**

One of the advantages of L-DED is its capability to rapidly fabricate large components at a high deposition rate [5, 52]. Meanwhile, it is well-suited for making functionally graded [8, 53] and heterostructured materials [6] for improved mechanical properties, due to its fine control over the microstructure and materials being deposited. In this work, we set up a multi-objective optimization problem to solve the conflict between these two objectives, aiming at the rapid fabrication of heterostructured alloys with low porosity. The optimization problem can be expressed as:

$$\mathbf{min}\ pt(P, v, \dot{m}, hs), res\ (P, v, \dot{m}, hs) \qquad (3)$$
$$sbj.t.:$$

$$P \in [30, 60] \qquad (4)$$
$$v \in [4, 12] \qquad (5)$$
$$\dot{m} \in [0.2, 0.8] \qquad (6)$$

$$hs \in [0.5, 0.7] \tag{7}$$
$$0.2 < ds(P, v, \dot{m}, hs) < 0.5 \tag{8}$$

Where $P$, $v$, $\dot{m}$, and $hs$ are the laser power, scan speed, powder feed rate, and hatch spacing, respectively. The upper and lower limits of $P$, $v$, and $\dot{m}$ are picked so that neither poor bonding nor high porosity was produced. $Pt$ is the print time for a cubic print with a fixed size of 10 mm×10 mm×5 mm (L × W × H), *res* the resolution of the prints, defined as the minimal resultant track width in a print (Figure 4), and $ds$ the dilution of multi-track prints, defined using the following equation:

$$ds = \frac{Multi\ track\ A_{remelt}}{Multi\ track\ A_{remelt} + Multi\ track\ A_{clad}} \tag{9}$$

Where $A_{clad}$ and $A_{remelt}$ are the areas of clad and remelt, respectively (Figure *2* (b) and (c)). The values of dilutions are constrained between 0.2 and 0.5, for strong bonding and minimal porosity [14, 54].

Figure 5 (c) depicts the detailed GA optimization loop. Before the start of this loop, two optimization objectives were defined, i.e., faster print speed and finer print resolution. Within each optimization generation, GA proposes solutions, i.e., the combinations of $P$, $v$, $\dot{m}$, and $hs$ that will be used as input to the two previously trained ML models to predict the corresponding single-track melt pools and multi-track prints, respectively. From the predicted multi-track prints the $t$ can be recovered for reconstructing fully dense multi-layer cubes. The optimization objectives are then calculated and compared for optimality. This loop continues until the optimization is converged.

The optimizations are set up using NSGA-iii, an advanced algorithm for finding Pareto optimal solutions [55, 56]. Pymoo, an open-source Python implementation for genetic algorithm (GA) [57] is used for performing the optimizations. Hypervolume [58] is calculated and used as an indicator of optimization convergency. Within each optimization generation, the objective functions are evaluated 1024 times with different solutions proposed by GA. Optimization is repeated three times initialized with different random seeds to ensure the influence of potential randomness is removed.

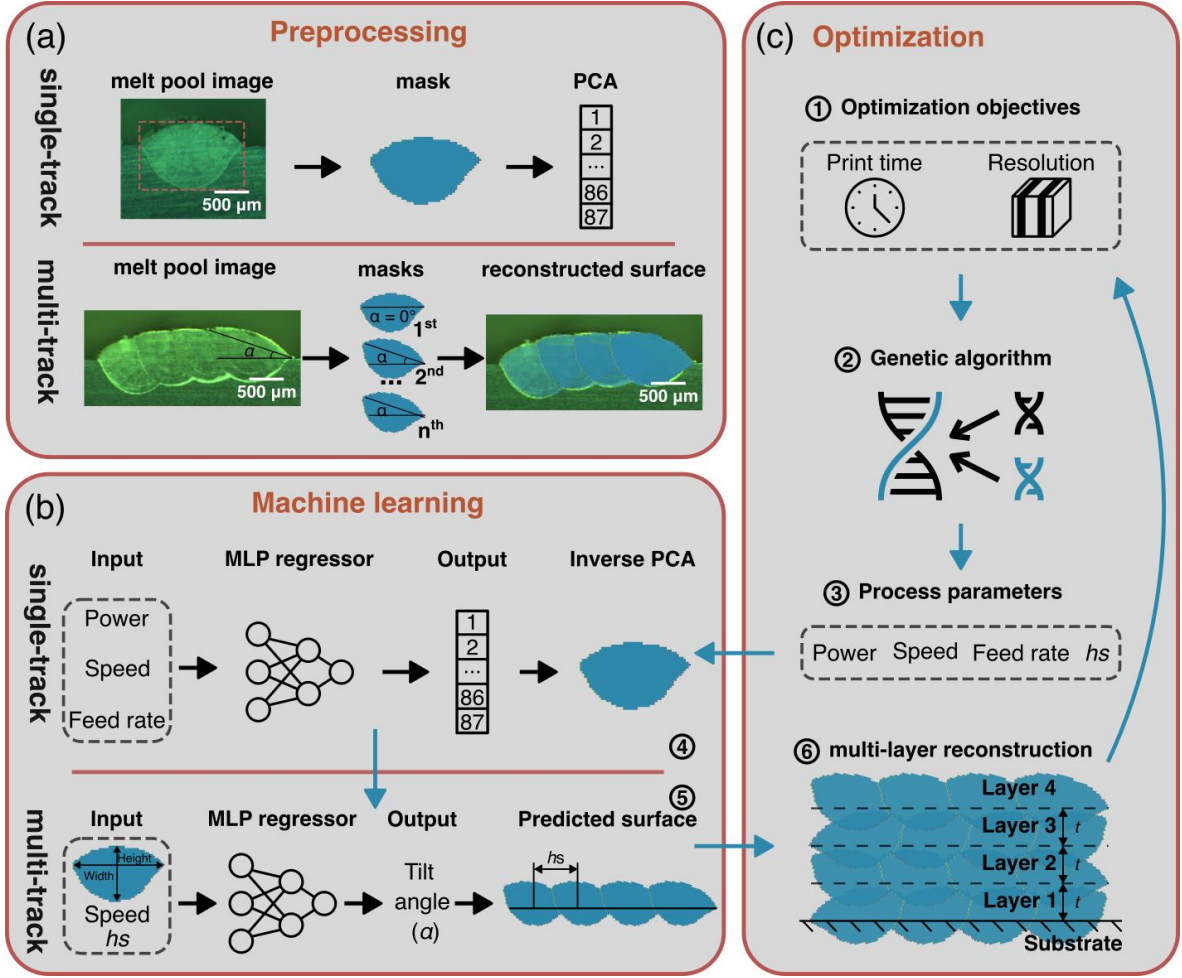

**Figure 5.** Machine learning and optimization methodology. (a) Data pre-processing for single-track (top) and multi-track (bottom) data generation. For single-track prints, melt pool contours are extracted and transformed into vectors using principal component analysis (PCA). Multi-track prints can then be reconstructed by tilting the first melt pool with an angle $\alpha$ and equally spaced at $hs$. (b) Machine learning inference models for single-track (top) and multi-track (bottom) prediction. Shown are the input parameters for each model and their corresponding outputs. (c) Genetic algorithm multi-objective optimization loop. With the print time (speed) and resolution as objectives, the genetic algorithm proposes new process parameters for the prediction and reconstruction of single-track, multi-track, and multi-layer prints. The objectives are evaluated at the end of each optimization loop until the optima are found.

## 3. Results

### 3.1. Single-track melt pools

In this work, we generated a comprehensive single-track melt pool dataset across a large design space. The influence of the basic process parameters, i.e., $P$, $v$, and $\dot{m}$, on the melt pool size and geometry is first investigated. Figure 6 gives the instances of single-track melt pool microscopic images at various combinations of these three basic process parameters. Each melt pool can be divided into two parts, namely a clad area formed by newly deposited material, and a remelt area (Figure 2 (b)). The clad area is a critical factor in the build rate, and the remelt area is usually used as an important indicator of intertrack bonding and porosity [14, 16, 54, 59].

As evident in Figure 6, melt pool size and geometry are largely dependent on the basic process parameters. As a general trend, the increase in $P$, decrease in $v$, and increase in $\dot{m}$ all lead to the increase in the overall melt pool size. Nevertheless, these three parameters influence the melt pool shapes in different ways. When $P$ varies while fixing the other two parameters, melt pool size variation is mainly from the change of the remelt area because of the change in energy input while the amount of newly deposited materials remains similar. When only $\dot{m}$ increases, the amount of newly deposited material changes while the energy input stays the same, causing a larger clad but smaller remelt area. In the case of changing $v$ solely, a balanced change in clad and remelt is found. This is due to that slower $v$ provides more energy input and gives more time for the material to melt and be deposited. In terms of melt pool geometry, it varies significantly across the process parameter space. A more symmetrical and regular melt pool geometry is obtained when the clad area is small and has a small curvature. In contrast, the melt pool geometry shows more irregularity as the clad gets larger. Overall, the melt pool size and geometry are complex signatures that are influenced by a combination of all basic parameters rather than any single one. Such complexity renders it advantageous to use ML methods for modeling the melt pool.

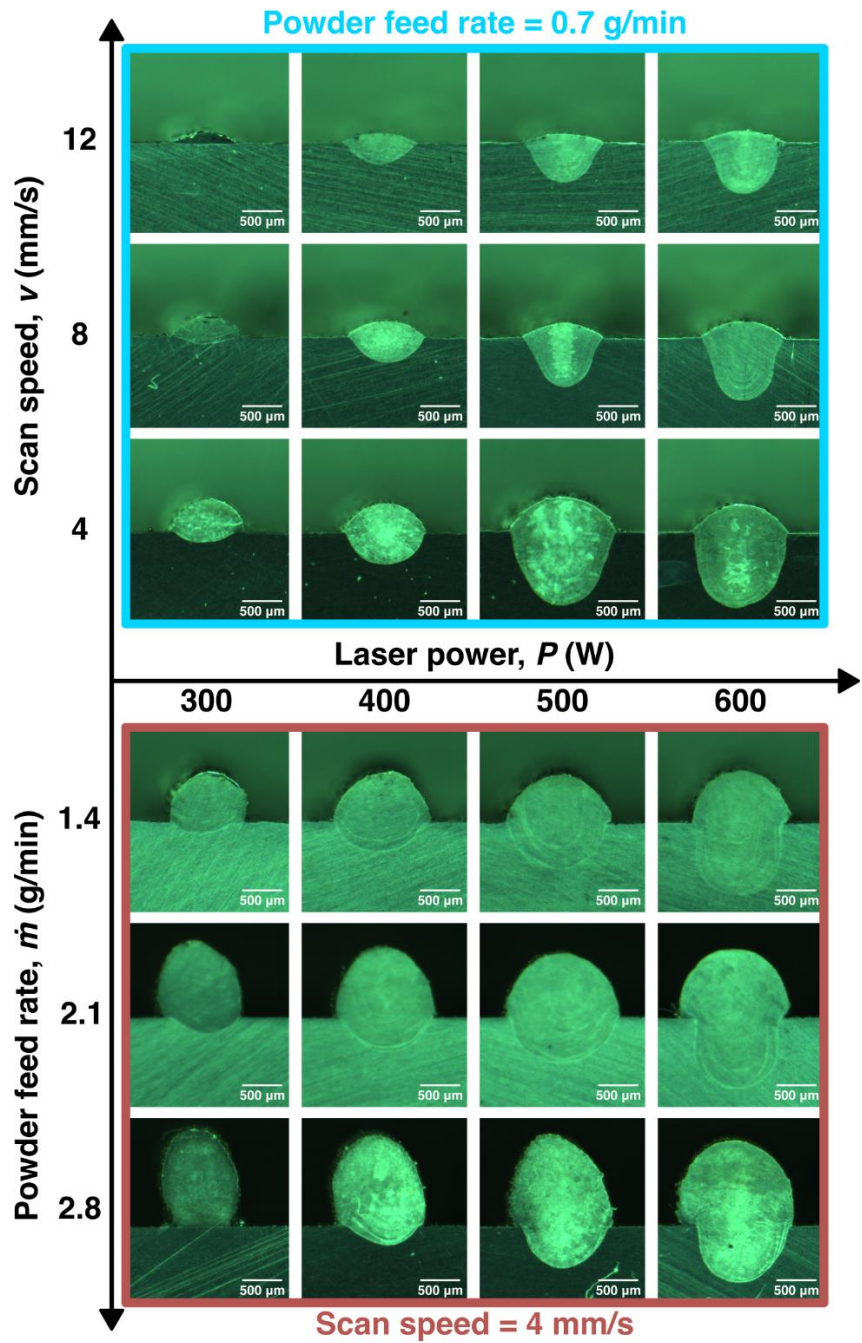

**Figure 6.** Single-track melt pool optical microscopic images and melt pool variations with the basic process parameters: laser power (*P*), scan speed (*v*), and powder feed rate (*ṁ*). The top rows show melt pools with a fixed *ṁ* and varying *P* and *v*, while the bottom rows show melt pools with a fixed *v* and varying *P* and *ṁ*.

### 3.3. Multi-track prints

Figure 7 shows multi tracks printed at combinations of *P*, *v*, and *hs*. Similar to single-track melt pools, the cross-sections of multi-track prints are divided into a clad area and a remelt area (Figure *2* (c)). The height of the clad area is defined as the layer height, a measurement of the height of newly added materials.

Figure 7 shows the size and geometry dependency of multi-track prints on the process parameters. Apart from the inherited trends similar to how $P$ and $v$ influence the single-track melt pools, the effect of $hs$ plays a dominant role in the geometry of multi-track prints. The smaller the $hs$ is, the more overlapping area there is between neighbouring tracks. When $hs$ is too small, i.e., $hs$ is 0.4, material accumulates and grows mostly along the height direction, making the remelt area penetrate the substrate less and thus poor bonding with the substrate. As $hs$ increases, the neighbouring tracks are more spread out. For $hs$ from 0.5 to 0.7, newly deposited tracks after the second or third track becomes a horizontal linear translation of the previous one. Thus, the multi-track prints can be estimated as equally spaced melt pools tilted at $α$, as explained previously in Section 2.3.1 (Figure 5 (a)). When $hs$ is too large, the resultant top contour will become less even, increasing the likelihood of lack-of-fusion pores when building a multi-layer part. When the $hs$ is greater than the single-track width, two neighbouring tracks will no longer be in contact and will fail to form an intact surface. Accordingly, in this work, we select $hs$ values ranging between 0.5 and 0.7 as printable parameters to use in the multi-objective optimization.

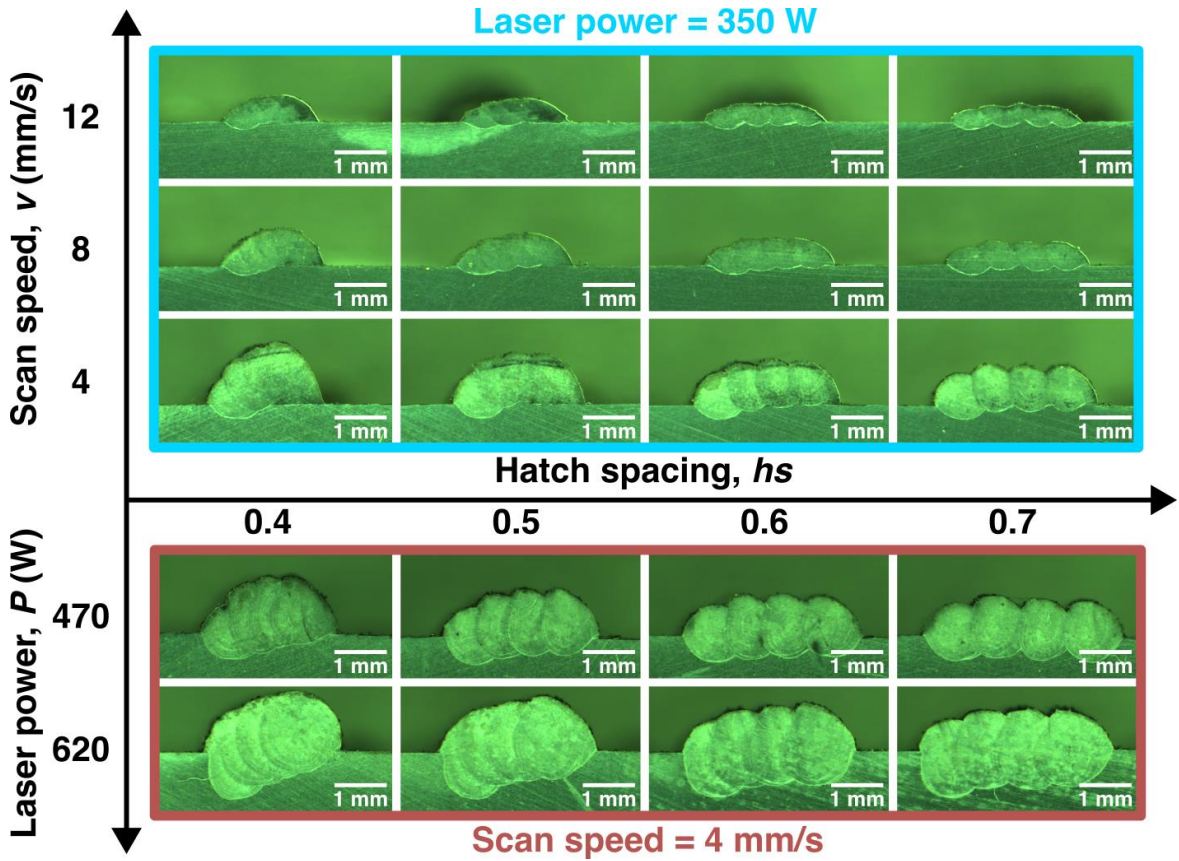

**Figure 7.** Multi-track melt pool optical microscopic images and melt pool variations with process parameters: laser power ($P$), scan speed ($v$), and hatch spacing ($hs$). The top rows show images with a fixed P and varying v and hs, while the bottom row*s* show images with a fixed v and varying P and hs.

## 3.4. ML model for single-track samples

Figure 8 (a) gives the single-track melt pool size distribution by showing the areas of the single-track prints collected in this work. In general, melt pool size increases with the increase in laser power and the decrease in scan speed, due to the increase in LED. Meanwhile, the melt pool size increases with the increase in powder feed rate as a result of more materials being deposited. Overall, a diversely distributed dataset is acquired for training the single-track ML model (Figure 8 (b)), with most melt pool areas around 0.5 mm$^2$. The smallest and largest melt pools are 0.06 mm$^2$ and 1.77 mm$^2$, respectively, and their process parameters are identified at the two extreme corners, i.e., minimum and maximum LEDs and feed rates (Figure 8 (c)).

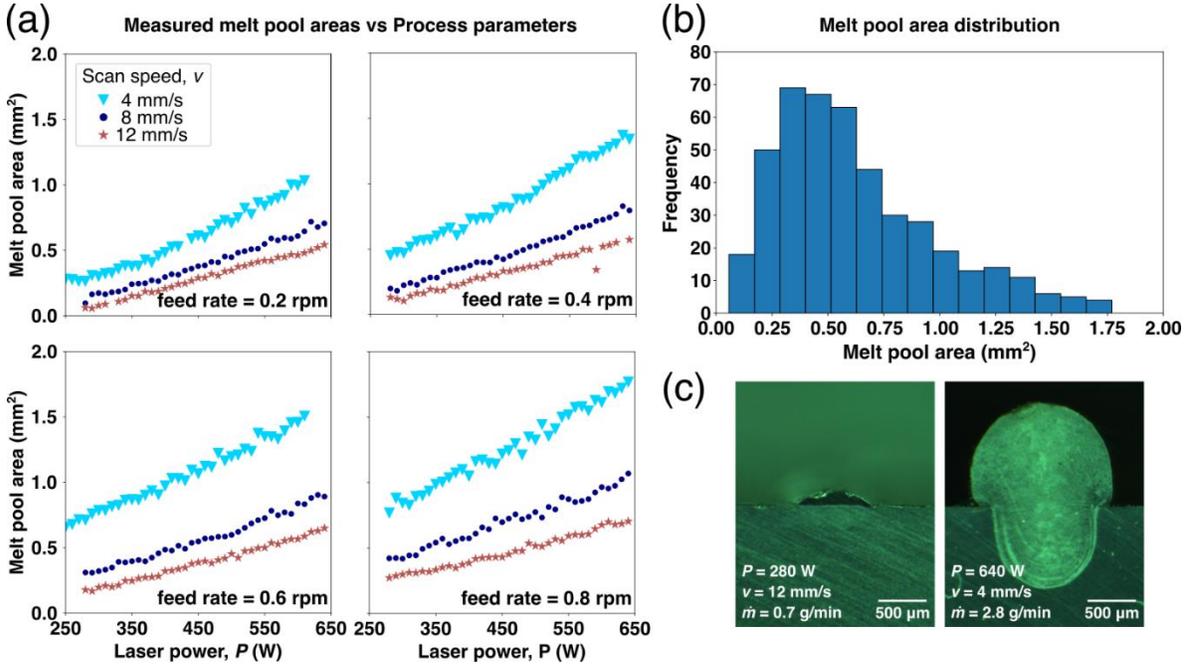

**Figure 8.** Single-track dataset statistics and optical microscopic images for two extreme cases. (a) Scattered plots showing the measured melt pool areas and their variations with the basic process parameters: laser power ($P$), scan speed ($v$), and powder feed rate ($\dot{m}$). (b) Histogram of the melt pool area distribution. (c) Two melt pools found at the extreme ends of the design space. On the left is the melt pool under the lowest $P$, $\dot{m}$, and the highest $v$; On the right is the melt pool under the highest $P$, $\dot{m}$, and the lowest $v$.

The single-track ML model was successfully trained to achieve a $R^2$ score of 0.995 on the stand-alone test set, for the MLP model when the melt pool area is compared for evaluation (Figure 9 (a)). The performances of the other two ML models, i.e., LR and RF, can be found in Figures S5(a) and (c). The MLP model and the RF model show similar performance, and both outperform the LR model. Although the MLP model is only slightly better than the RF model, it is selected in this work due to its capability of performing transfer learning, for the ease of transferring our framework to other material systems. Furthermore, since the goal of the single-track model is not only to accurately predict the melt pool size but also the exact melt pool geometry, in Figure 9 (b) (and Figures S5 (b) and (d) for the RF and LR models, respectively) we give comparisons between microscopic and ML-predicted melt pool contours across a wide range of process parameters. The ML model accurately captures the melt pool contours using $P$, $v$, and $\dot{m}$ as input, with an inference time of around 10 milliseconds, which further enables the setup of an accurate and efficient optimization of the framework.

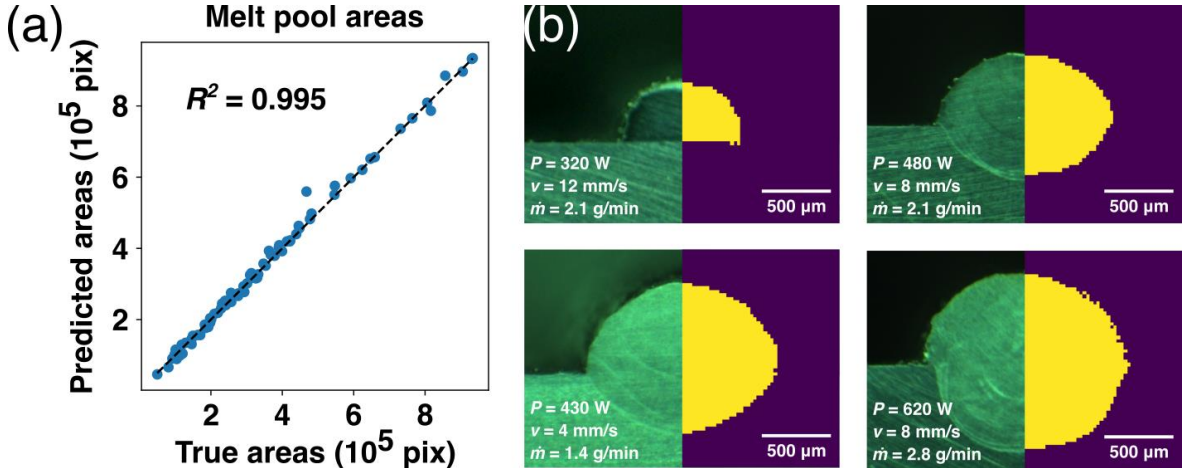

**Figure 9.** Single-track ML model performance. (a) A parity plot comparing the true melt pool areas and the predicted melt pool areas. (b) Comparisons between melt pool contours from microscopic images (left), and their corresponding ML predicted contours (right).

### 3.5. Multi-track ML model

Figure 10 (a) gives the measured tilt angles ($\alpha$) in the multi-track dataset. As a general trend, an increase in $hs$ would result in a decrease in $\alpha$ because of the larger spacing between the neighbouring tracks. Meanwhile, the increase in $v$ leads to a smaller $\alpha$ since it reduces the clad height of previous tracks (Figure 7). It is interesting to note that despite the increase in $P$ leads to an increase in LED, the resulting $\alpha$ shows a slight decrease. This is because when $\dot{m}$ and $v$ are fixed, the amount of powder being deposited remains roughly the same even though the LED increases. Consequently, the increase in $P$ causes the melt pool to penetrate the previous tracks more, resulting in a smaller $\alpha$ (Figure 7). Figure 10 (b) shows the overall $\alpha$ distribution, which shows a close to normally distributed distribution with a mean of 27.15° and a standard deviation of 8.51°.

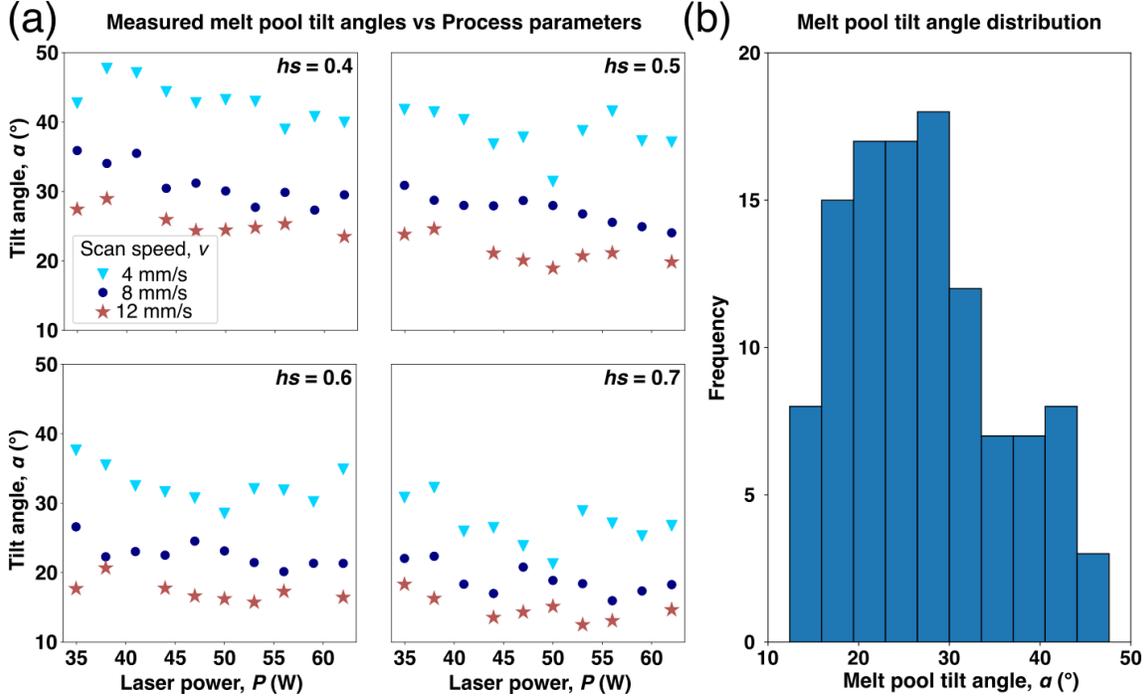

**Figure 10.** Multi-track dataset statistics. (a) Scattered plots showing the measured melt pool tilt angles ($α$) and their variations with laser power ($P$), scan speed ($v$), and hatch spacing ($hs$). (b) Histogram of the melt pool tilt angle distribution for all samples.

The multi-track ML model was trained to achieve a $R^2$ score of 0.969 on the stand-alone test set using the MLP model (Figure 11 (a)), indicating that using the single-track melt pool $H$ and $W$, process parameters $v$ and $hs$ as input, $α$ can be accurately predicted. Combining $α$ with the predicted melt pool mask from the previously trained single-track model, the predicted multi-track can be reconstructed with a good agreement to real-life optical microscopic images (Figure 11 (b), where the real-life melt pool contours are marked with solid black lines and the ML predictions are marked with dashed white lines). Despite such satisfactory matching, errors exist in matching exactly the reconstructed multi-track prints with the real-life case. Notably, the right-most melt pool of the true print extends further compared with the reconstructed predictions (Figure 11 (b)), due to the deposited materials being caught on the printing substrate surface. Such differences may result in inaccuracy when predicting the width of the printed multi-track, but only to a limited extent since it becomes less pronounced as the number of tracks in a single layer increases. Lastly, it should be noted that in this work, the extrapolation capability of the ML models, i.e., their capability to make predictions outside of the design space used in this work, was not tested. This is because those parameters not within the selected design space either result in inadequate or excessive energy input, rendering them difficult to print.

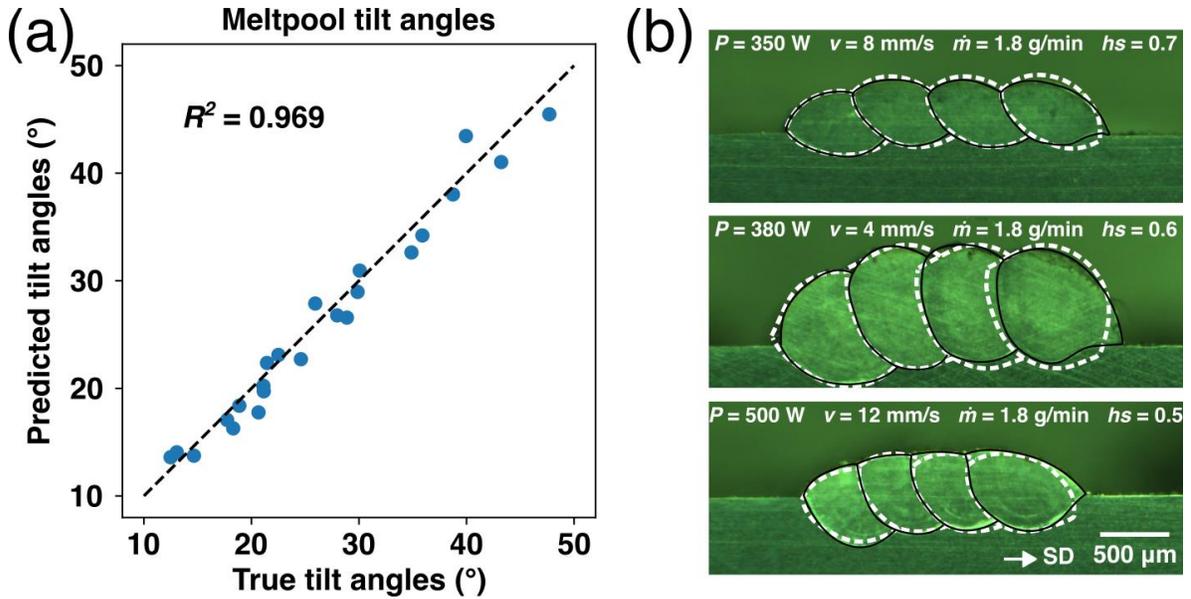

**Figure 11.** Multi-track ML model performance. (a) A parity plot comparing the true and predicted melt pool tilt angles ($\alpha$) (b) Comparisons between melt pool contours from microscopic images (solid black lines) and their corresponding predicted contours (dashed white lines).

### 3.6. Multi-layer reconstruction

Furthermore, we demonstrate the performance of our framework in reconstructing multi-layer prints. Figure 12 shows the microscopic images of the cross-sections of three multi-layer prints produced under different process parameters. Outlined in dashed white lines are the reconstructed melt pool contours by our framework. On predicting the overall cross sections of the prints (Figure 12 (a)-(c)), although the width along the scan direction (SD) is being captured accurately, we see a deviation in the height along the build direction (BD). Moreover, the top corners of the real-life prints are rounded compared with those from the framework reconstruction. This phenomenon is more pronounced in the high laser power and high powder feed rate cases as shown in Figure 12 (a) and (b). However, when comparing locally the melt pools at areas close to the centre-bottom of the builds, the predicted contours match well with the real-life prints (Figure 12 (d)-(f)).

In the prediction of single-track melt pools, the melt pool shape is mainly a result of the complex thermal history during the printing process. When it comes to multi-track and multi-layer, the geometrical factors, i.e., tilting and horizontally patterning in the multi-track case, and vertically patterning in the multi-layer case (Figure 5 (b) and (c)), play a more significant role. Thus, as long as the melt pool shape and tilt angle are accurately predicted, there will be a good agreement between the real-life prints and reconstructed multi-track and multi-layer prints (Figure 11 and Figure 12). However, despite the accuracy of the two ML models, errors exist and will accumulate, eventually leading to more significant mismatches along the BD for multi-layer predictions. Moreover, when the tracks are deposited in real life, the first melt pool being deposited in a multi-track print is never tilted, causing the starting end of the multi-track to be lower in height (Figure 7). As a result, the corners of the multi-layer prints are observed to be lower and "rounded". Furthermore, smaller melt pools generally have smaller tilt angles (Figure 7) and thus smaller differences in height within the same layer, causing the error to be less pronounced for builds with smaller single-track melt pools (Figure 12 (a)-(c)).

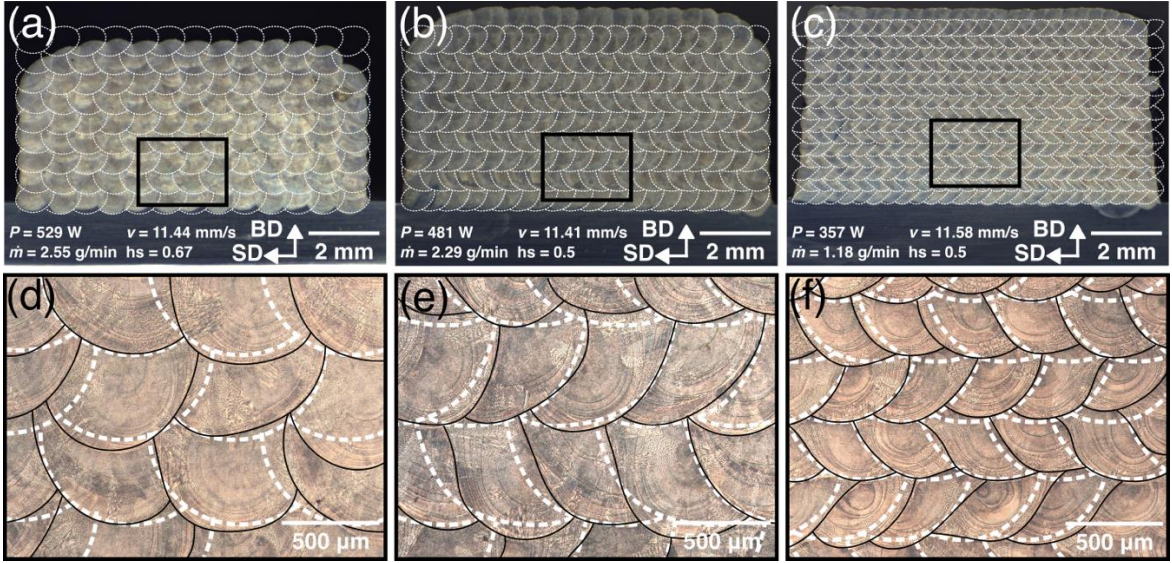

**Figure 12.** Comparisons between real-life and AIDED reconstructed multi-layer prints under various printing conditions. (a)-(c) Overall comparisons of the cross sections of three 10 mm × 10 mm × 5 mm (L × W × H) cubic prints. Reconstrued melt pool contours are marked with dashed white lines. (d)-(f) Zoomed in local comparison between melt pool arrangements in the locations indicated in (a)-(c). AIDED reconstructed contours are marked with dashed white lines and real-life melt pool contours are marked with solid black lines.

**3.7. Framework transferability**

To demonstrate the high transferability of our framework, we show results of a new single-track ML model trained with transfer learning for pure Ni (Figure 13 (a) and (b)). Despite the limited amount of training data, i.e., 56 single-tracks, this model achieved a $R^2$ score of 0.902 on the stand-alone test set, when melt pool areas are compared (Figure 13 (a)). The exact melt pool contours are then compared in Figure 13 (b), showing a good agreement between real-life melt pools and ML predictions. Furthermore, a comparison between the performances of the transfer-learned model and a model trained without transfer learning is provided in Figure 13 (c) and (d). This time, the melt pool area prediction $R^2$ score is 0.827, with poorer performance in predicting exact melt pool contours too. These results highlight the transferability of our model to accurately predict melt pool characteristics across different material systems even when trained with a limited amount of data, leveraging the benefits of transfer learning.

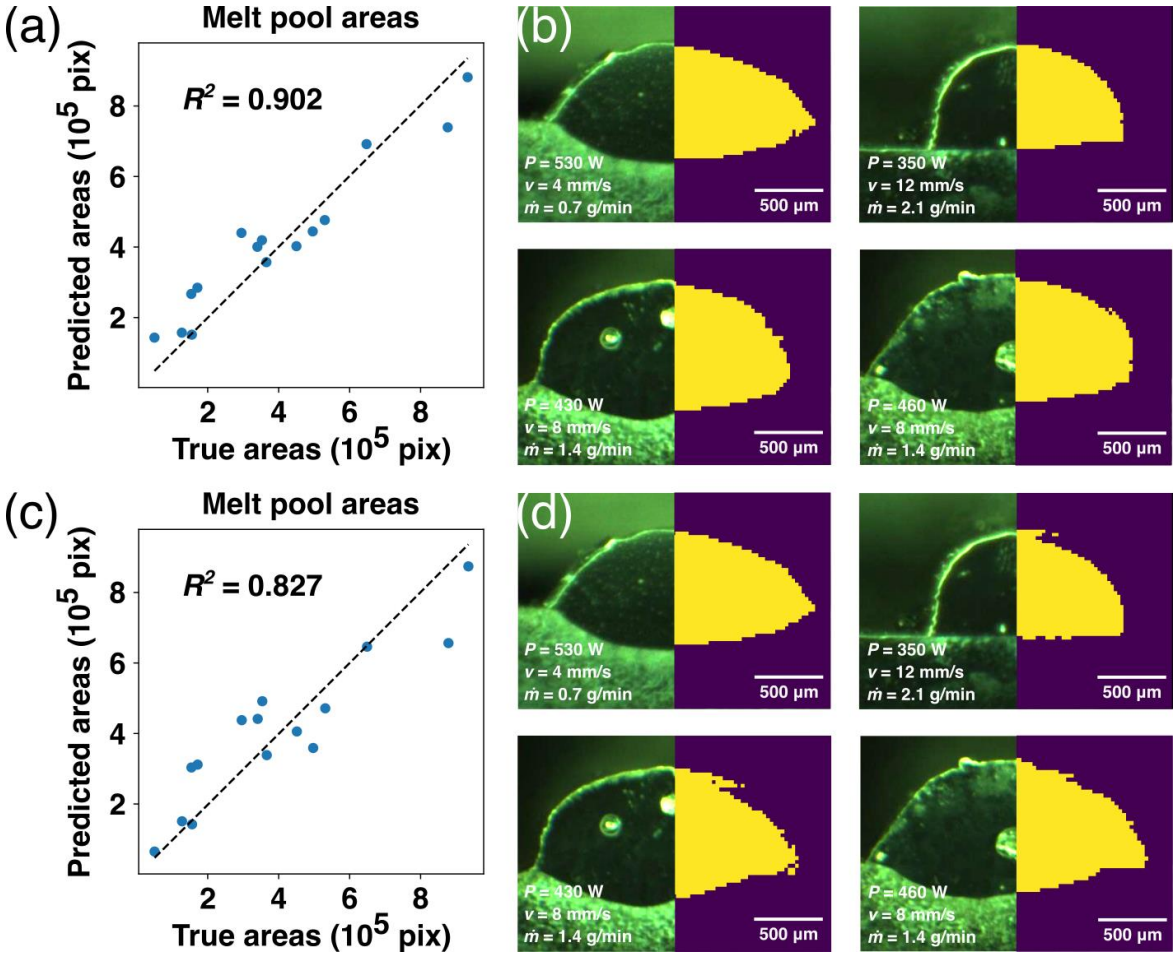

**Figure 13.** Comparison between single-track models with and without transfer learning on the Ni dataset. (a) and (c) Parity plots comparing the true melt pool areas and the predicted melt pool areas. (b) and (d) Comparisons between melt pool contours from microscopic images (left), and their corresponding ML predicted contours (right).

### 3.8. Multi-objective optimization

Figure 14 shows a summary of the multi-objective optimization results. For all three optimization repetitions, the simulation converged within 15 generations (Figure 14 (a)). A Pareto front is identified for both optimization objectives in Figure 14 (b). From the detailed optimization results (Table 3) we show that the finest resolution of the L-DED system used in this work, despite having a 1 mm diameter laser beam, can reach as fine as 0.392 mm. A comparison between prints with a fine resolution and a coarse one can be found in Figure 4, demonstrated using heterostructured steel cubes printed under the optima from the optimization results. For the finest resolution and fastest print speed cases, the resolutions, measured as the smallest width of the material bands, are 0.392 mm and 0.725 mm, respectively. As far as the print time is concerned, for a 10 mm × 10 mm x 5 mm cube the fastest is 101.9 seconds, a 62.6% reduction from when the finest resolution is achieved.

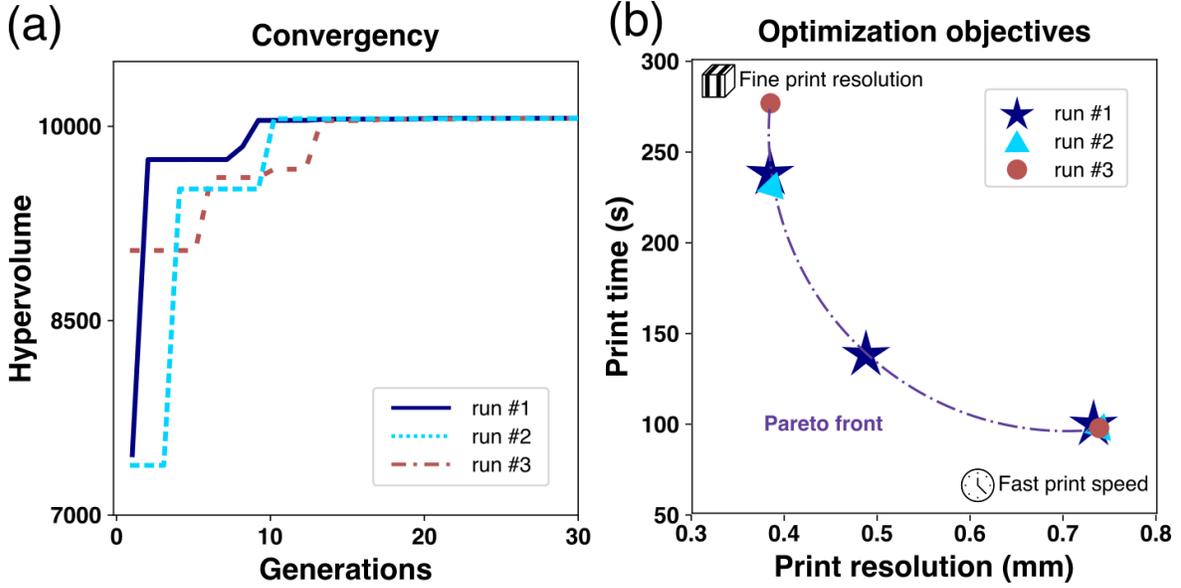

**Figure 14.** Multi-objective optimization results. (a) Convergency plot for the three optimization runs performed in this work. (b) Global optima for the two competing optimization objectives, i.e., print time and resolution. A Pareto front is identified by a dashed line, along where the optima from each optimization run reside.

**Table 3**

Detailed multi-objective optimization results showing the optima from the three optimization runs.

| Sample | Optimal process parameters | | | | Optimization objectives | |
|---|---|---|---|---|---|---|
| | $P$ (W) | $V$ (mm/s) | $\dot{m}$ (g/min) | hs | Resolution (mm) | Print time (s) |
| 1-1 | 543 | 11.42 | 2.43 | 0.67 | 0.725 | 102.5 |
| 1-2 | 481 | 11.61 | 2.29 | 0.50 | 0.490 | 139.5 |
| 1-3 | 357 | 11.58 | 1.18 | 0.50 | 0.392 | 233.1 |
| 2-1 | 526 | 11.49 | 2.58 | 0.67 | 0.730 | 101.9 |
| 2-2 | 361 | 10.58 | 0.71 | 0.50 | 0.392 | 272.2 |
| 3-1 | 529 | 11.44 | 2.55 | 0.67 | 0.730 | 102.3 |
| 3-2 | 357 | 11.84 | 1.29 | 0.50 | 0.392 | 228.0 |

To validate the dimensional accuracy and density for the optimal prints, we provide detailed measurements of the optimal results in Table 4 and Figure 15. In terms of the cube width ($W$) (Figure 15 (a)), all prints achieved the accuracy with an average error of 1.07% with a standard deviation of 0.87% from the designed width. For the inherently more difficult cube height ($H$) (Figure 15 (b)), the results show the accuracy with an average error of 10.76% with a standard deviation of 3.67% from the designed heights. The cube length ($L$) is not considered here as it is consistently accurate across all cubes and is a dimension irrelevant to the process parameter selection under the printing path used in this work (Figure 3). Notably, as shown in Figure 15 (c), all printed cubes are nearly fully dense, showing densities all above 99.9%. Their average density is 99.980% with a standard deviation of 0.018%. For each sample, we measured its densities at three distinct locations, i.e., top, middle, and bottom (Figure 15 (c)). The density values shown in Table 4 are averages of the densities at the three locations for each print.

**Table 4**

Dimensions and density measurements of the identified optima

| Sample | Cube W (mm) | | error (%) | Cube H (mm) | | error (%) | Density (%) |
|---|---|---|---|---|---|---|---|
| | designed | actual | | designed | actual | | |
| **1-1** | 10 | 9.72 | -2.8 | 4.898 | 4.45 | -9.15 | 99.847 |
| **1-2** | 10 | 10.1 | 1 | 4.898 | 5.47 | 11.68 | 99.941 |
| **1-3** | 10 | 10 | 0 | 4.898 | 5.31 | 8.41 | 99.978 |
| **2-1** | 10 | 9.89 | -1.1 | 4.898 | 4.62 | -5.68 | 99.988 |
| **2-2** | 10 | 9.95 | -0.5 | 4.877 | 4.07 | -16.55 | 99.989 |
| **3-1** | 10 | 9.96 | -0.4 | 4.898 | 4.48 | -8.53 | 99.984 |
| **3-2** | 10 | 10.17 | 1.7 | 4.898 | 5.65 | 15.35 | 99.985 |
| **mean** | | | 1.07 | | | 10.76 | 99.959 |
| **std** | | | 0.87 | | | 3.67 | 0.052 |

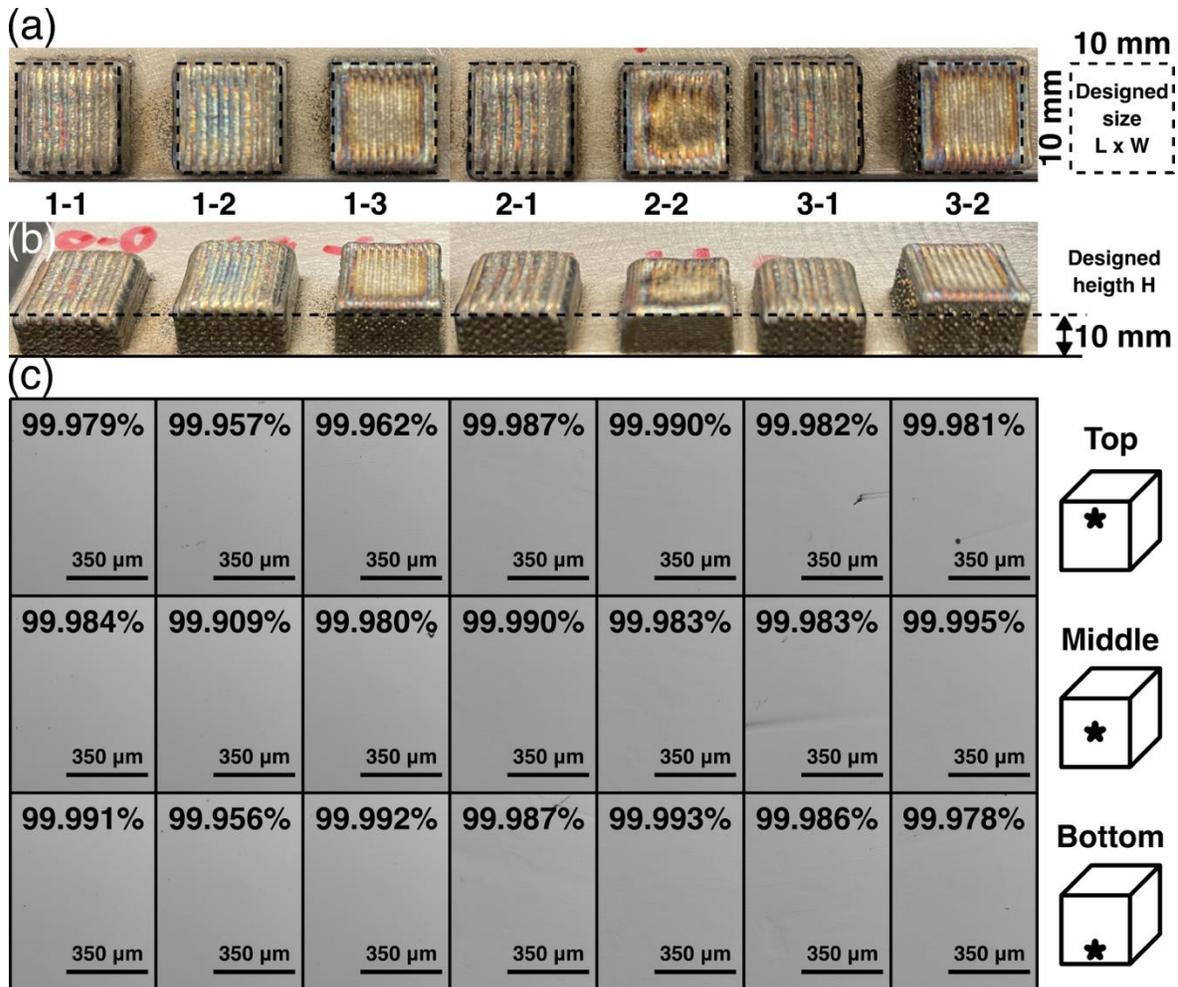

**Figure 15.** Dimension and density measurements of the optimal prints from all optimization runs. (a) Top view of the prints. Marked in dashed black lines are the designed dimensions. (b) Trimetric view of the prints. Marked in the dashed black line is the designed height, i.e., 10 mm. (c) Measured densities of each print at the top, middle, and bottom locations.

Finally, we generated a series of processing maps from the developed ML models, showing the variations of the two objectives, i.e., print time and print resolution, in the multi-objective optimization along with the melt pool dilution (Figure 16). Each plot was produced against a combination of two process parameters from $P$, $v$, $\dot{m}$, and $hs$, while the two other process parameters are set at the average of their selected processing range. An optimal is marked by a red star for each plot in Figure 16, aiming for a balance between the objectives. Guided by these processing maps, designers could easily identify the optimal process values according to their design purposes.

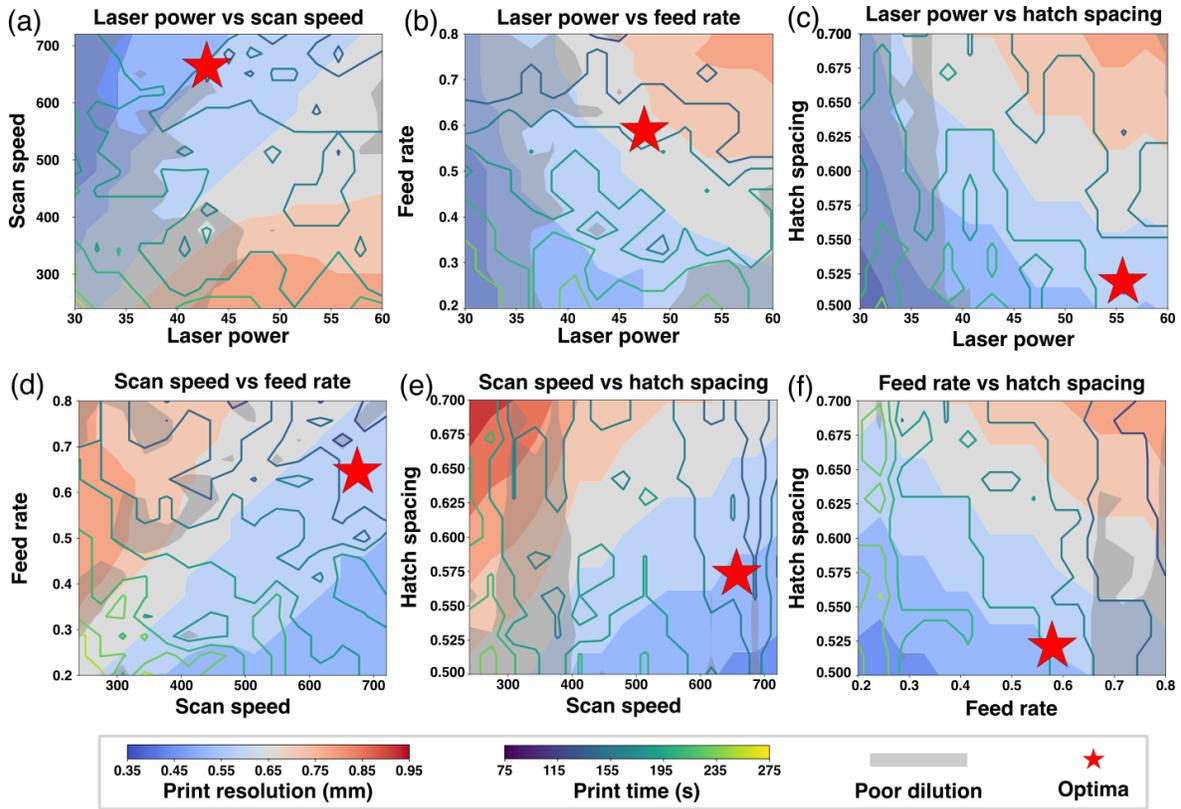

**Figure 16.** Processing maps showing the variations of print resolution, time, and dilution as functions of different combinations of process parameters. The optima are highlighted with a red star for each map. On these processing maps, the background colours show the print resolution, the isolines show print time, and the grey shade indicates poor dilution.

## 4. Discussion

The dataset generated in this work covers a wide range of process parameters in the design space for 316L stainless steel and is by far the largest dataset for melt pool prediction in the L-DED process to the best of the authors knowledge. The dominating influence on the melt pool size and geometry by four processing parameters, i.e., $P$, $v$, $\dot{m}$, and $hs$, is observed and discussed. The complex nature of how the combinations of these process parameters influence the melt pool size and geometry justifies the advantages of using ML-based techniques for the prediction of melt pool characteristics. The two well-trained ML models show applaudable accuracy in capturing the exact size and geometry of single-track and multi-track melt pools, yielding $R^2$ scores of 0.995 and 0.966, respectively. Such results exceed those in previous studies [35, 42], where more complex processing and material features were used. We credit this improvement to the direct use of exact melt pool contours, which preserves the completest possible melt pool geometrical information during the model training stage. The accuracy and swiftness of our models further lays the foundation for accurate multi-layer reconstruction and efficient process optimization.

For multi-layer prints, low prediction errors averaging 1.07% and 10.76% are achieved for the width (*W*) and height (*H*) of cubic prints, respectively. The prediction of *H*, a parameter dependent on the selection of layer thickness, has been studied in various previous works and the prediction errors are all over 20% for cubes of 5 mm height [14, 60], rendering our work an over 46.2% improvement without the use of extra online sensing hardware, which are complex and costly. Furthermore, all multi-layer prints using optimal parameters from AIDED have densities greater than 99.9% and can be considered as fully dense cubes.

Benefiting from the accurate and swift inference of our ML models, AIDED offers inverse optimization capabilities for various industrial objectives. In this work, we use two objectives to demonstrate how AIDED aids the inverse selection of optimal process parameters to rapidly create heterostructured steels with coarse and fine print resolutions. The optimization process can be finished in as soon as one to three hours, compared with traditional methods, which are known to be time-consuming [61, 62]. Moreover, one can easily set up other optimization objectives in AIDED based on their own needs, following the steps detailed in this paper. Owing to the good transferability of the MLP models used in this work, with a small amount of new data one can transfer ADED for other printable materials. Here we demonstrate that uisng less than 60 new single tracks, we can successfully transfer AIDED from 316L stainless steel to pure nickel.

Nevertheless, we recognize the limitations and room for improvements of AIDED, and some of the limitations and potential solutions are discussed here. First, despite the good transferability of ML models, the current AIDED is material-specific and not generalized. To improve, more data from a wider range of dissimilar materials should be collected, and intrinsic material properties should be included as inputs. Second, despite achieving good accuracy in multi-layer height prediction, there is still room for improvement as 10.76% error is still considerably large. To achieve this, thermal history information should be introduced into the model training and multi-layer print reconstruction stage. Last, the performance of AIDED for complex-shaped prints has not been studied. In future work, this should be investigated, and improvements should be made accordingly.

5. **Conclusions**

In this work, we have developed the AIDED framework – an *Accurate Inverse* process optimization framework in L-*DED* using machine learning and optimization approaches. This framework offers high accuracy and efficiency in the prediction of melt pool size and geometry in the L-DED process. Moreover, AIDED rapidly finds optimal process parameters directly from customizable optimization objectives. The high transferability of this framework makes it versatile to be applied for a large variety of materials. More specifically, AIDED contributes to the current L-DED field in the following ways:

1) It allows researchers to accurately predict single-track, multi-track, and multi-layer melt pool size and geometry, which is beneficial for replacing time-consuming experiments to generate large datasets for studying complex physics in L-DED. Predictions show the accuracies of 0.995 and 0.966 for single-track and multi-track prints, respectively, when the $R^2$ score is calculated. The prediction errors of the width and height of multi-layer prints are 1.07% and 10.76%, respectively.

2) This framework allows designers to set up optimizations with tailored objectives so that optimal process parameters can be inversely identified with high efficiency. Owing to the swift inference time of the machine learning models, the optimization typically completes in 1-3 hours. The optimality of the results is experimentally validated aiming at high print resolution and fast print speed. When a 10 mm × 10 mm × 5 mm ($L \times W \times H$) multi-layer cube is considered, the resolution is as fine as 0.392 mm using a 1 mm diameter laser beam, and the shortest print time is 101.9 s, a 62.6% reduction from the highest resolution case.

3) The framework is highly transferrable: Using less than 60 single-track melt pool data, AIDED can be transferred to be used in a new material, i.e., pure nickel.

4) The processing maps generated by AIDED can serve as a starting point for machine calibration and the selection of process window at a low cost and short time.

5) The detailed steps described in this work allows researchers to create their versions of AIDED to "aid" various research and industrial applications efficiently.

**Acknowledgements**

This work was supported by the Natural Sciences and Engineering Research Council of Canada (NSERC) Discovery Grant (RGPIN-2018-05731), Centre for Analytics and Artificial Intelligence Engineering (CARTE) Seed Funding program, New Frontiers in Research Fund-Exploration (NFRFE-2019-00603), the Data Sciences Institute Catalyst Grant, and the Connaught Innovation Award from the University of Toronto. X.S. acknowledges the financial support from the NSERC Canada Graduate Doctoral Scholarship (CGS-D). We thank Ethan Wang Jan for his help in processing the samples in a machine shop.

**Author contributions: CRediT**

**Xiao Shang:** Conceptualization, Methodology, Software, Validation, Data Curation, Visualization, Writing – original draft, Writing – review & editing. **Evelyn Li:** Software, Validation, Data Curation, Visualization, Writing – review & editing. **Ajay Talbot:** Validation, Data Curation, Writing – review & editing. **Haitao Wen**: Validation, Data Curation, Visualization, Writing – review & editing. **Tianyi Lyu:** Validation, Writing – review & editing. **Jiahui Zhang**: Validation, Writing – review & editing. **Yu Zou:** Supervision, Funding, Writing – review & editing.

**Data availability**

The complete framework and the dataset used has been released in an open-source GitHub repository at: https://github.com/xshang93/AIDED

**Declaration of Competing Interest**

The authors declare that they have no known competing financial interests or personal relationships that could have appeared to influence the work reported in this paper.

**References**

1. Gu, D., et al., *Material-structure-performance integrated laser-metal additive manufacturing.* Science, 2021. **372**(6545): p. eabg1487.
2. DebRoy, T., et al., *Scientific, technological and economic issues in metal printing and their solutions.* Nature Materials, 2019. **18**(10): p. 1026-1032.
3. Herzog, D., et al., *Additive manufacturing of metals.* Acta Materialia, 2016. **117**: p. 371-392.
4. Poudel, A., et al., *Feature-based volumetric defect classification in metal additive manufacturing.* Nature Communications, 2022. **13**(1): p. 6369.
5. Svetlizky, D., et al., *Directed energy deposition (DED) additive manufacturing: Physical characteristics, defects, challenges and applications.* Materials Today, 2021. **49**: p. 271-295.
6. Tan, C., et al., *Additive manufacturing of voxelized heterostructured materials with hierarchical phases.* Additive Manufacturing, 2022. **54**: p. 102775.
7. Zhang, C., et al., *Additive manufacturing of functionally graded materials: A review.* Materials Science and Engineering: A, 2019. **764**: p. 138209.
8. Zhu, Z., et al., *Recent progress on the additive manufacturing of aluminum alloys and aluminum matrix composites: Microstructure, properties, and applications.* International Journal of Machine Tools and Manufacture, 2023. **190**: p. 104047.
9. Piscopo, G. and L. Iuliano, *Current research and industrial application of laser powder directed energy deposition.* The International Journal of Advanced Manufacturing Technology, 2022. **119**(11): p. 6893-6917.
10. Gasser, A., et al., *Laser Additive Manufacturing.* Laser Technik Journal, 2010. **7**(2): p. 58-63.
11. Najmon, J.C., S. Raeisi, and A. Tovar, *2 - Review of additive manufacturing technologies and applications in the aerospace industry*, in *Additive Manufacturing for the Aerospace Industry*, F. Froes and R. Boyer, Editors. 2019, Elsevier. p. 7-31.
12. Kumar, L.J. and C.G. Krishnadas Nair, *Current Trends of Additive Manufacturing in the Aerospace Industry*, in *Advances in 3D Printing & Additive Manufacturing Technologies*, D.I. Wimpenny, P.M. Pandey, and L.J. Kumar, Editors. 2017, Springer Singapore: Singapore. p. 39-54.
13. Zhang, K., et al., *Pore evolution mechanisms during directed energy deposition additive manufacturing.* Nature Communications, 2024. **15**(1): p. 1715.
14. Vaughan, M.W., et al., *Development of a process optimization framework for fabricating fully dense advanced high strength steels using laser directed energy deposition.* Additive Manufacturing, 2023. **67**: p. 103489.
15. Nenadl, O., et al., *A versatile model for the prediction of complex geometry in 3D direct laser deposition.* Surface and Coatings Technology, 2016. **307**: p. 292-300.
16. Bax, B., et al., *Systematic evaluation of process parameter maps for laser cladding and directed energy deposition.* Additive Manufacturing, 2018. **21**: p. 487-494.
17. Kürnsteiner, P., et al., *High-strength Damascus steel by additive manufacturing.* Nature, 2020. **582**(7813): p. 515-519.
18. Adomako, N.K., et al., *Microstructures and mechanical properties of multi-layered materials composed of Ti-6Al-4V, vanadium, and 17–4PH stainless steel produced by directed energy deposition.* Additive Manufacturing, 2022. **59**: p. 103174.
29

19. Zhu, Y., et al., *Ultrastrong nanotwinned titanium alloys through additive manufacturing.* Nature Materials, 2022. **21**(11): p. 1258-1262.
20. Zhou, Z., et al., *Effects of process parameters on microstructure and cracking susceptibility of a single crystal superalloy fabricated by directed energy deposition.* Materials & Design, 2021. **198**: p. 109296.
21. Mazzucato, F., et al., *Influence of Process Parameters and Deposition Strategy on Laser Metal Deposition of 316L Powder.* Metals, 2019. **9**(11): p. 1160.
22. Zhang, W., et al. *A Layer-Dependent Analytical Model for Printability Assessment of Additive Manufacturing Copper/Steel Multi-Material Components by Directed Energy Deposition*. Micromachines, 2021. **12**, DOI: 10.3390/mi12111394.
23. Traxel, K.D., et al., *Model-driven directed-energy-deposition process workflow incorporating powder flowrate as key parameter.* Manufacturing Letters, 2020. **25**: p. 88-92.
24. Ramiro, P., et al., *Strategy Development for the Manufacturing of Multilayered Structures of Variable Thickness of Ni-Based Alloy 718 by Powder-Fed Directed Energy Deposition.* Metals, 2020. **10**(10): p. 1280.
25. Liu, F.Q., et al., *On the varieties of build features during multi-layer laser directed energy deposition.* Additive Manufacturing, 2020. **36**: p. 101491.
26. Wei, H.L., et al., *Prediction of spatiotemporal variations of deposit profiles and inter-track voids during laser directed energy deposition.* Additive Manufacturing, 2020. **34**: p. 101219.
27. Caiazzo, F. and V. Alfieri, *Simulation of Laser-assisted Directed Energy Deposition of Aluminum Powder: Prediction of Geometry and Temperature Evolution.* Materials, 2019. **12**(13): p. 2100.
28. Toyserkani, E. and A. Khajepour, *A mechatronics approach to laser powder deposition process.* Mechatronics, 2006. **16**(10): p. 631-641.
29. Hu, D. and R. Kovacevic, *Sensing, modeling and control for laser-based additive manufacturing.* International Journal of Machine Tools and Manufacture, 2003. **43**(1): p. 51-60.
30. Everton, S.K., et al., *Review of in-situ process monitoring and in-situ metrology for metal additive manufacturing.* Materials & Design, 2016. **95**: p. 431-445.
31. Smoqi, Z., et al., *Closed-loop control of meltpool temperature in directed energy deposition.* Materials & Design, 2022. **215**: p. 110508.
32. Behlau, F., et al., *Layer thickness controlling in Direct Energy Deposition process by adjusting the powder flow rate.* Procedia CIRP, 2022. **111**: p. 330-334.
33. Borovkov, H., et al., *In-Line Height Measurement Technique for Directed Energy Deposition Processes.* Journal of Manufacturing and Materials Processing, 2021. **5**(3): p. 85.
34. Johnson, N.S., et al., *Invited review: Machine learning for materials developments in metals additive manufacturing.* Additive Manufacturing, 2020. **36**: p. 101641.
35. Akbari, P., et al., *MeltpoolNet: Melt pool characteristic prediction in Metal Additive Manufacturing using machine learning.* Additive Manufacturing, 2022. **55**: p. 102817.
36. Meng, L., et al., *Machine Learning in Additive Manufacturing: A Review.* JOM, 2020. **72**(6): p. 2363-2377.
37. Bayat, M., et al., *A review of multi-scale and multi-physics simulations of metal

*additive manufacturing processes with focus on modeling strategies.* Additive Manufacturing, 2021. **47**: p. 102278.

38. Criales, L.E., et al., *Laser powder bed fusion of nickel alloy 625: Experimental investigations of effects of process parameters on melt pool size and shape with spatter analysis.* International Journal of Machine Tools and Manufacture, 2017. **121**: p. 22-36.
39. Keshavarzkermani, A., et al., *An investigation into the effect of process parameters on melt pool geometry, cell spacing, and grain refinement during laser powder bed fusion.* Optics & Laser Technology, 2019. **116**: p. 83-91.
40. Yu, G., et al., *Influence of processing parameters on laser penetration depth and melting/re-melting densification during selective laser melting of aluminum alloy.* Applied Physics A, 2016. **122**(10): p. 891.
41. Dilip, J.J.S., et al., *Influence of processing parameters on the evolution of melt pool, porosity, and microstructures in Ti-6Al-4V alloy parts fabricated by selective laser melting.* Progress in Additive Manufacturing, 2017. **2**(3): p. 157-167.
42. Feenstra, D.R., A. Molotnikov, and N. Birbilis, *Utilisation of artificial neural networks to rationalise processing windows in directed energy deposition applications.* Materials & Design, 2021. **198**: p. 109342.
43. Seede, R., et al., *An ultra-high strength martensitic steel fabricated using selective laser melting additive manufacturing: Densification, microstructure, and mechanical properties.* Acta Materialia, 2020. **186**: p. 199-214.
44. Seede, R., et al., *Effect of composition and phase diagram features on printability and microstructure in laser powder bed fusion: Development and comparison of processing maps across alloy systems.* Additive Manufacturing, 2021. **47**: p. 102258.
45. Shi, K., et al., *Process-structure multi-objective inverse optimisation for additive manufacturing of lattice structures using a physics-enhanced data-driven method.* Virtual and Physical Prototyping, 2023. **18**(1): p. e2266641.
46. Wang, Z., et al., *A Data-Driven Approach for Process Optimization of Metallic Additive Manufacturing Under Uncertainty.* Journal of Manufacturing Science and Engineering, 2019. **141**(8).
47. Aboutaleb, A.M., et al., *Multi-Objective Accelerated Process Optimization of Part Geometric Accuracy in Additive Manufacturing.* Journal of Manufacturing Science and Engineering, 2017. **139**(10).
48. Jocher, G., A. Chaurasia, and J. Qiu, *YOLO by Ultralytics.* 2023.
49. Kirillov, A., et al. *Segment anything.* in *Proceedings of the IEEE/CVF International Conference on Computer Vision*. 2023.
50. Pedregosa, F., et al., *Scikit-learn: Machine learning in Python.* the Journal of machine Learning research, 2011. **12**: p. 2825-2830.
51. Li, Z., Y.P. Lai, and E. Diller, *3D Printing of Multilayer Magnetic Miniature Soft Robots with Programmable Magnetization.* Advanced Intelligent Systems. **n/a**(n/a): p. 2300052.
52. Ahn, D.-G., *Directed Energy Deposition (DED) Process: State of the Art.* International Journal of Precision Engineering and Manufacturing-Green Technology, 2021. **8**(2): p. 703-742.
53. Ji, W., et al., *Recent progress in gradient-structured metals and alloys.* Progress in

Materials Science, 2023. **140**: p. 101194.
54. Bhardwaj, T., et al., *Direct Energy Deposition - Laser Additive Manufacturing of Titanium-Molybdenum alloy: Parametric studies, microstructure and mechanical properties.* Journal of Alloys and Compounds, 2019. **787**: p. 1238-1248.
55. Jain, H. and K. Deb, *An Evolutionary Many-Objective Optimization Algorithm Using Reference-Point Based Nondominated Sorting Approach, Part II: Handling Constraints and Extending to an Adaptive Approach.* IEEE Transactions on Evolutionary Computation, 2014. **18**(4): p. 602-622.
56. Deb, K. and H. Jain, *An Evolutionary Many-Objective Optimization Algorithm Using Reference-Point-Based Nondominated Sorting Approach, Part I: Solving Problems With Box Constraints.* IEEE Transactions on Evolutionary Computation, 2014. **18**(4): p. 577-601.
57. Blank, J. and K. Deb, *Pymoo: Multi-Objective Optimization in Python.* IEEE Access, 2020. **8**: p. 89497-89509.
58. Fonseca, C.M., L. Paquete, and M. Lopez-Ibanez. *An Improved Dimension-Sweep Algorithm for the Hypervolume Indicator*. in *2006 IEEE International Conference on Evolutionary Computation*. 2006.
59. de Oliveira, U., V. Ocelík, and J.T.M. De Hosson, *Analysis of coaxial laser cladding processing conditions.* Surface and Coatings Technology, 2005. **197**(2): p. 127-136.
60. Shim, D.-S., et al., *Effect of layer thickness setting on deposition characteristics in direct energy deposition (DED) process.* Optics & Laser Technology, 2016. **86**: p. 69-78.
61. Era, I.Z., et al., *Machine learning in Directed Energy Deposition (DED) additive manufacturing: A state-of-the-art review.* Manufacturing Letters, 2023. **35**: p. 689-700.
62. Bikas, H., M.A. Terzakis, and P. Stavropoulos *Manufacturability-Based Design Optimization for Directed Energy Deposition Processes*. Machines, 2023. **11**, DOI: 10.3390/machines11090879.